# High-pressure torsion processing of serine and glutamic acid: Understanding mechanochemical behavior of amino acids under astronomical impacts

*Dedicated to Professor Reinhard Pippan, a great scientist in the field of high-pressure torsion, on the occasion of his 70th birthday*

Kaveh Edalati[1,2,*], Jacqueline Hidalgo-Jiménez[1,2], Thanh Tam Nguyen[1,3], Motonori Watanabe[1], and Ikuo Taniguchi[4]

[1] WPI, International Institute for Carbon-Neutral Energy Research (WPI-I2CNER), Kyushu University, Fukuoka 819-0395, Japan
[2] Graduate School of Integrated Frontier Sciences, Department of Automotive Science, Kyushu University, Fukuoka 819-0395, Japan
[3] Mitsui Chemicals, Inc. - Carbon Neutral Research Center (MCI-CNRC), Kyushu University, Fukuoka 819-0395, Japan
[4] Faculty of Fiber Science and Technology, Kyoto Institute of Technology, Kyoto 606-8585, Japan

Astronomical impacts by small solar system bodies (meteoroids, asteroids, comets, and transitional objects) are considered a mechanism for delivering amino acids and their polymerization to proteins in early Earth conditions. High-pressure torsion (HPT) is a new methodology to simulate such impacts and clarify the behavior of biomolecules. In this study, two amino acids, crystalline *L*-serine and *L*-glutamic acid that were detected in meteorites, are processed by HPT and examined by ex situ X-ray diffraction, Raman spectroscopy, nuclear magnetic resonance, Fourier transform infrared spectroscopy, and in situ mechanical shear testing. No polymerization, chemical reactions, or phase transformations are detected after HPT, indicating that the stability and presence of these two amino acids in meteorites are quite reasonable. However, some microstructural and mechanical changes like crystal size reduction to the nanometer level, crystal defect formation, lattice expansion by vacancy formation, and shear strength enhancement to the steady state are found which are similar to the behaviors reported in metals and ceramics after HPT processing.

*Corresponding author (E-mail: kaveh.edalati@kyudai.jp; Tel/Fax: +81 92 802 6744)

**Introduction**

One main question regarding the origin of life on the early Earth is the mechanism of formation of proteins and ribonucleic acid (RNA), as two essential polymers for life [1]. Amino acids, which are monomers for the formation of proteins, were detected in a few meteorites [2-4]. This discovery has led to a suggestion that small solar system bodies such as meteoroids, asteroids, comets, and transitional objects could deliver these essential biomolecules to the young Earth four billion years ago [5]. There are also attempts to prove that impacts by these small solar system bodies could lead to the polymerization of amino acids to form the first proteins [6].

High-pressure shock experiments are generally used to simulate the behavior of biomolecules under astronomical impacts [7-9], but these shock experiments ignore the plastic strain effect which could be significant when small solar system bodies collide with the Earth's crust. The high-pressure torsion (HPT) process is a new methodology in this field that can simulate such impacts [10]. In the HPT method, which was originally introduced by Bridgman, not only the effect of high pressure but also the effect of strain can be included in the simulations [11,12]. Details of HPT processes were explained in a few review papers [13-15], but the method can be simply explained as the application of torsional strain under high pressure to a material between two massive anvils. Although HPT has been used to synthesize big molecules of metals [16-18] and ceramics (oxides [19], oxynitrides [20], and hydrides [21]) from the mixture of elements and was occasionally employed in the past for processing organic materials [22-24], its application to biomolecules to form bigger molecules is quite limited [10].

In the first application of HPT as a tool for the simulation of astronomical impacts, an amino acid, glycine ($C_2H_5NO_2$), was examined [10]. Glycine as the simplest amino acid, was selected because of its detection in meteorites [2-4] and comets [9,25]. HPT processing of glycine did not lead to polymerization but decomposed it to alcohol and nitrogen-containing molecules [10]. This study suggested that the effect of astronomical impacts on mechanochemical reactions on the Earth could be quite intense in the presence of plastic strain. Despite this report, the behavior of other amino acid biomolecules under such high-pressure and strain conditions is not well understood. Since besides glycine, other amino acids like serine and glutamic acid were also detected in meteorites [2-4], their behavior under impacts needs to be investigated by HPT processing. Serine has a composition of $C_3H_7NO_3$ and crystallizes as an orthorhombic phase with the space group of $P2_12_12_1$, and glutamic acid has a composition of $C_5H_9NO_4$ and crystallizes as orthorhombic structures with the space group of $P2_12_12_1$ in two forms of the α and β phases with different lattice parameters [26,27]. Both serine and glutamic acid are non-essential amino acids and can be synthesized by the human body, and their *L*-type isomers are naturally found in proteins.

In this study, *L*-serine and *L*-glutamic acid are processed by the HPT method and their mechanochemical behaviors are studied by different characterization methods. It is found that these two amino acids are quite stable under high pressure and strain in good agreement with their presence in meteorites [2-4]. Moreover, the microstructural refinement and mechanical property enhancement of these two biomolecules appear to be similar to those reported for metals [28,29] and ceramics [30] processed by HPT.

**Experimental Materials and Procedures**

Powders of *L*-serine and *L*-glutamic acid (the β phase) with purity levels of 99% were prepared from Fujifilm Wako, Japan. About 100 mg of powders were compressed using a manual hydraulic press under 380 MPa to produce discs with 10 mm diameter for HPT processing. The HPT facility had two anvils: the upper anvil was fixed in a rigid metallic frame, while the lower anvil could move up and down using a hydraulic press and rotate using an electric motor. The anvils had cylindrical shape (50 mm diameter and 50 mm height), made of a composite of tungsten carbide and 11 wt% cobalt, and had a circular flat-bottom hole (10 mm diameter and 0.25 mm depth) on the surface. The discs were placed on the hole of the lower anvil, the lower anvil moved up to touch the upper anvil with a force of 470 kN (48 tons) which is equivalent to a pressure of 6 GP. The discs were processed under 6 GPa at either room temperature or 423 K, and shear strain was introduced by the rotation of the lower HPT anvil against the upper anvil for 5 turns with a rotation speed of one turn per minute. The compressive load, pressure and rotation speed remained constant and uniform during the process. The samples after HPT processing had a disc shape with 10 mm diameter, but their thickness was smaller than the thickness of the initial samples. While the initial discs were ~1 mm thick, their thickness decreased to ~0.7 mm after HPT processing. Further details about the HPT process were reported in earlier publications [13-15]. The HPT-processed discs and initial powders were examined by various characterization techniques, as described below.

The crystal structure of samples was examined by ex situ X-ray diffraction (XRD) using Cu Kα radiation (λ: 1.5418 Å). The lattice parameters and lattice volume of the samples were determined from the XRD profiles using a PDXL software equipped with the Rietveld method. Moreover, the crystallite size of the samples was determined using the Williamson-Hall [31] and Halder-Wagner [32] methods.

Phase transformation was further examined using ex situ Raman spectroscopy (Renishaw inVia Raman microscope) with laser wavelengths of 532 and 785 nm.

Information on chemical bonds was collected at room temperature by ex situ attenuated total reflectance-Fourier transform infrared (ATR-FTIR) spectroscopy equipped with a prism.

The polymerization and chemical reactions were examined by ex situ $^1$H and $^{13}$C nuclear magnetic resonance (NMR) spectra that were recorded on AVANCE III (Bruker, 600 MHz). About 20 mg of samples were dissolved in 0.5 mL of deuterium oxide (D$_2$O) for *L*-serine and a solution of 20 vol% of deuterium chloride (DCl) in D$_2$O for *L*-glutamic acid. The solutions were placed in NMR tubes and the resonance was measured at frequencies of 600 MHz and 150 MHz for $^1$H and $^{13}$C spectra, respectively.

The mechanical properties of samples were examined by in situ torque measurements. The shear strain (*γ*) and shear stress (*τ*) were estimated using the following equations [12].

$$\gamma = \frac{2\pi RN}{h} \tag{1}$$

$$\tau = \frac{3q}{2\pi R^3} \tag{2}$$

where *R* is the disc radii, *N* is the number of lower anvil rotations, *h* is the disc thickness and *q* is the in situ measured torque.

**Results**

Examination of crystal structures by XRD is summarized in Fig. 1 for (a) *L*-serine and (b) *L*-glutamic acid. Pristine *L*-serine and *L*-glutamic acid powders have orthorhombic crystal structures in agreement with JCPDF card numbers 00-030-1741 (*L*-serine I phase) and 00-054-2280 (β *L*-glutamic acid phase), respectively. Although pressure-induced phase transformations were reported earlier in both *L*-serine [33,34] and *L*-glutamic acid [35,36], no new peaks or indications for irreversible phase transformations appear in the XRD profiles. These results indicate that both materials remain stable under HPT processing conditions, which is in line with their survival in meteorites [2-4]. Two changes, however, can be observed in the XRD profiles:

peak shifts to lower diffraction angles and peak broadening, as shown more clearly in a magnified view of XRD profiles in Fig. 1c for *L*-serine and Fig. 1d for *L*-glutamic acid. Peak shift is an indication of lattice expansion due to the formation of point defects and vacancies [37], and peak broadening in crystalline materials is an indication of lattice strain due to the formation of dislocations and crystallite size reduction [38].

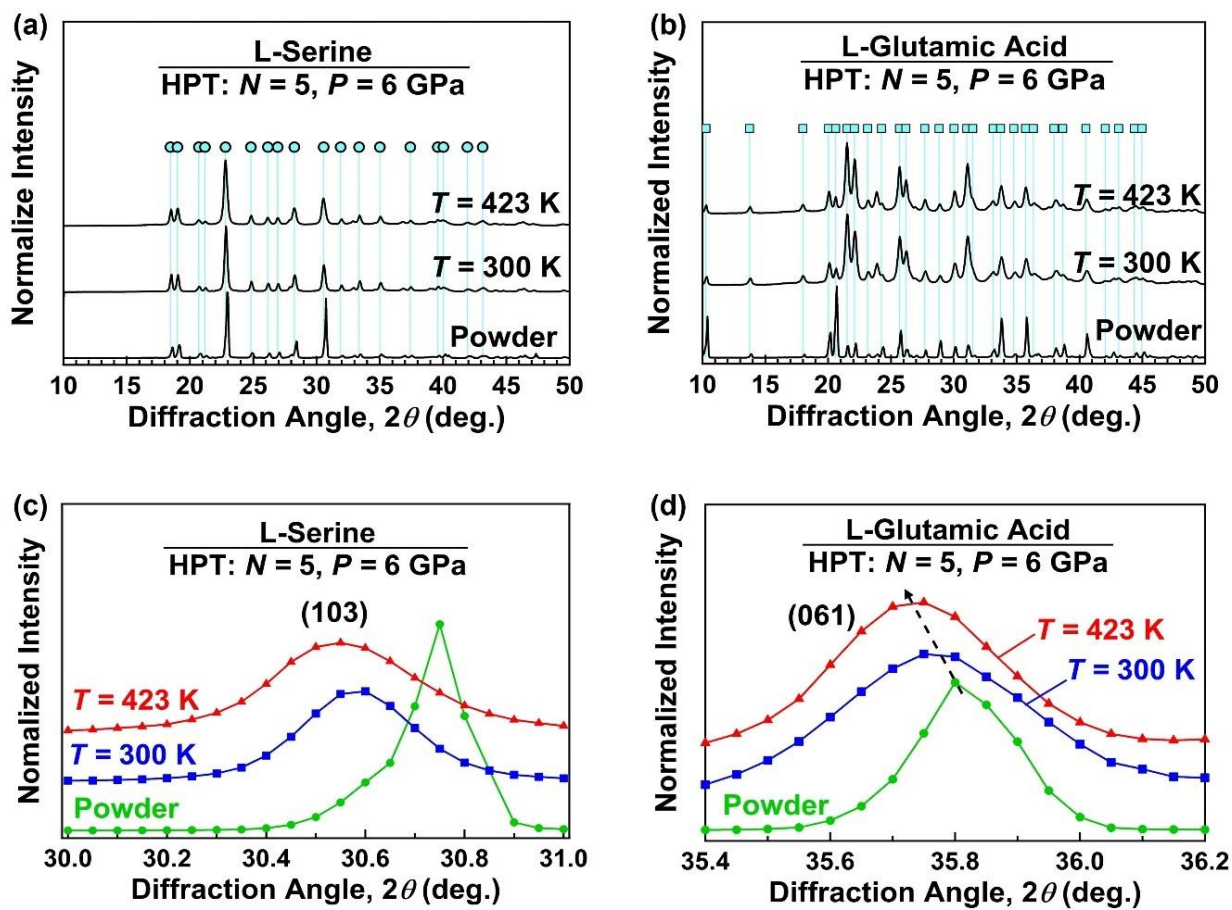

Figure 1. (a, b) Overall XRD patterns and (b) magnified patterns for (a, c) *L*-serine and (b, d) *L*-glutamic acid before and after HPT treatment at room temperature and 423 K.

Lattice parameters are given in Table 1 for amino acids before and after HPT processing and lattice expansions are plotted in Fig. 2a. It is found that lattice parameters increase and lattice volume expands after HPT processing for both amino acids, and these changes become more significant by increasing the temperature. Lattice expansion despite applying high pressure should be due to the formation of vacancies [37]. The formation of vacancies during HPT processing is a general phenomenon reported in both metallic materials [39,40] and ceramics [41]. An estimation of crystallite size using the Williamson-Hall [31] and Halder-Wagner [32] methods is shown in Fig. 2b and Fig. 2c, respectively. Since the validity of these two XRD line profile methods has not been confirmed in crystalline biomolecules by comparing them with microscopy results, the quantity of data in Fig. 2b and Fig. 2c should be evaluated by care. However, for both materials, crystallite size clearly decreases with the application of HPT. The reduction in crystallite size is enhanced in *L*-serine by increasing the processing temperature to 423 K, but it becomes less significant for *L*-glutamic acid at 423 K. The reduction in crystallite size after HPT processing is a general feature of almost all metallic materials [13,28,29] and ceramics [30].

Table 1. Crystal structure, space group and lattice parameters for *L*-serine and *L*-glutamic acid before and after HPT treatment at room temperature and 423 K.

|  | *L*-Serine | | | *L*-Glutamic Acid | | |
|---|---|---|---|---|---|---|
|  | Powder | 300K | 423K | Powder | 300K | 423K |
| Crystal Structure | Orthorhombic | | | Orthorhombic | | |
| Space Group | $P2_1 2_1 2_1$ | | | $P2_1 2_1 2_1$ | | |
| $a$ (Å) | 8.549 ± 0.009 | 8.553 ± 0.010 | 8.580 ± 0.030 | 6.911 ± 0.006 | 6.932 ± 0.007 | 6.938 ± 0.005 |
| $b$ (Å) | 5.581 ± 0.005 | 5.598 ± 0.007 | 5.605 ± 0.016 | 17.228 ± 0.011 | 17.243 ± 0.012 | 17.252 ± 0.008 |
| $c$ (Å) | 9.307 ± 0.007 | 9.327 ± 0.013 | 9.330 ± 0.020 | 5.141 ± 0.005 | 5.144 ± 0.006 | 5.149 ± 0.004 |

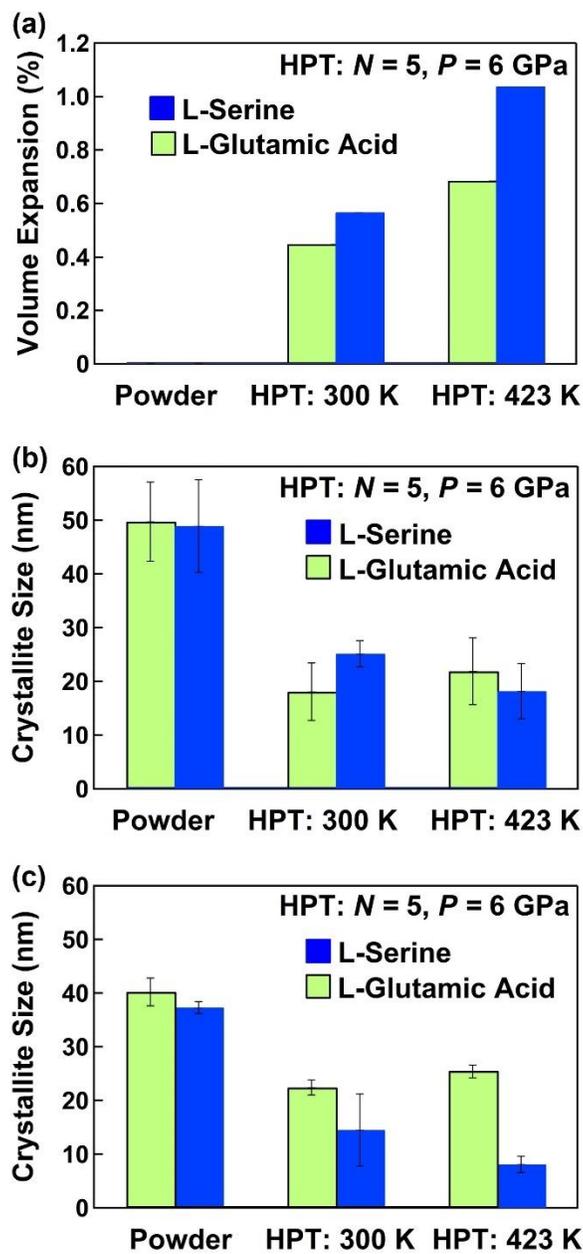

Figure 2. (a) Lattice volume expansion, (b) crystallite size estimated by Williamson-Hall method and (c) crystallite size estimated by Halder-Wagner method for *L*-serine and *L*-glutamic acid before and after HPT treatment at room temperature and 423 K.

Phase transformations were further examined by Raman spectroscopy, as shown in Fig. 3 for (a, c) *L*-serine and (b, d) *L*-glutamic acid using laser wavelengths of (a, b) 532 and (c, d) 785 nm. For the pristine powders, the Raman spectra are consistent with the reported spectra for orthorhombic *L*-serine [42-45] and the β orthorhombic phase of *L*-glutamic acid [35,42,44,46-48]. Table 2 shows the vibration modes for the detected peaks in both amino acids. It should be noted that no vibration modes have been assigned for some peaks in the literature. After HPT processing, the peak-to-noise ratio becomes weak, particularly when a high-energy laser with a 532 nm wavelength is used. The increase in the background intensity is due to the formation of lattice defects and particularly vacancies which was reported in some irradiated materials and amino acids [49,50]. Raman spectra do not provide a shred of clear evidence for the irreversible phase transformations or chemical reactions, in agreement with XRD analyses.

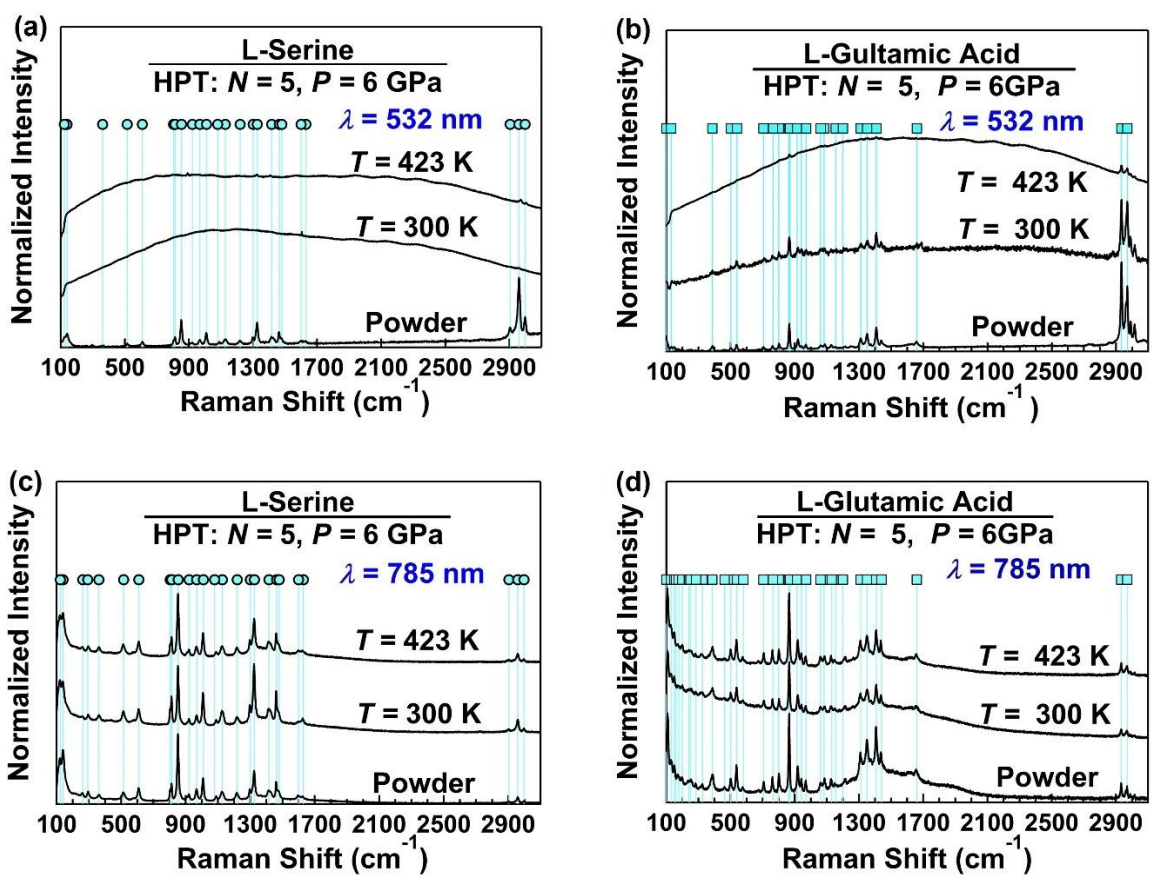

Figure 3. Raman spectra achieved using laser wavelengths of (a, b) 532 and (c, d) 785 nm for (a, c) *L*-serine and (b, d) *L*-glutamic acid before and after HPT treatment at room temperature and 423 K.

Table 2. Rama shifts and assigned vibration modes for *L*-serine and *L*-glutamic acid. Vibration modes were gathered from [35,42-48], while vibration modes were not reported for some Raman shifts.

| *L*-Serine | | *L*-Glutamic Acid | |
|---|---|---|---|
| Shift (cm$^{-1}$) | Vibration Mode | Shift (cm$^{-1}$) | Vibration Mode |
| 123 | - | 130 | Lattice |
| 140 | - | 150 | Lattice |
| 262 | - | 160 | Torsion |
| 293 | - | 178 | - |
| 362 | - | 200 | $CO_2^-$ torsion |
| 515 | - | 240 | - |
| 610 | $COO^-$ bending | 250 | - |
| 805 | - | 285 | Bending |
| 814 | $COO^-$ out-of-plane vibration | 325 | - |
| 854 | C-C stretching | 387 | B-polymorph lattice |
| 922 | - | 463 | - |
| 969 | C-N stretching | 502 | $CO_2^-$ bending |
| 1010 | C-O-H | 540 | $CO_2^-$ rocking |
| 1080 | C-N stretching | 580 | $CO_2^-$ |
| 1127 | - | 705 | B-polymorph lattice |
| 1220 | COH bending | 765 | $C-H_2$ rocking |
| 1301 | C-H rocking | 800 | B-polymorph C-C stretching |
| 1327 | C-H | 863 | B-polymorph |
| 1417 | COH bending | 871 | COOH |
| 1464 | $C-H_2$ rocking | 915 | C-C-N |
| 1480 | COH bending | 941 | B-polymorph C-C out-of-plane |
| 1600 | $COO^-$ stretching & $NH_3^+$ bending | 970 | C-C stretching |
| 1630 | $NH_3^+$ asymmetric bending | 1060 | B-polymorph C-N stretching |
| 2904 | C-H stretching | 1082 | C-O stretching |
| 2959 | $C-H_2$ symmetric stretching | 1130 | $NH_3^+$ rocking |
| 2998 | - | 1408 | B-polymorph |
| | | 1422 | $COO^-$ symmetric stretching |
| | | 1440 | - |
| | | 1660 | - |
| | | 2933 | - |
| | | 2970 | $C-H_2$ asymmetric stretching |

To examine chemical bonding, ATR-FTIR spectroscopy was used as shown in Fig. 4 for (a) *L*-serine and (b) *L*-glutamic acid. The profiles before and after HPT processing are reasonably similar (except for some small peak shifts) and consistent with the literature [51,52], indicating that no chemical reactions occur during HPT processing. Although temperature is an important processing parameter in HPT [12-15], ATR-FTIR spectra suggest that processing under a higher temperature of 423 K is also ineffective in initiating chemical reactions in these two amino acids.

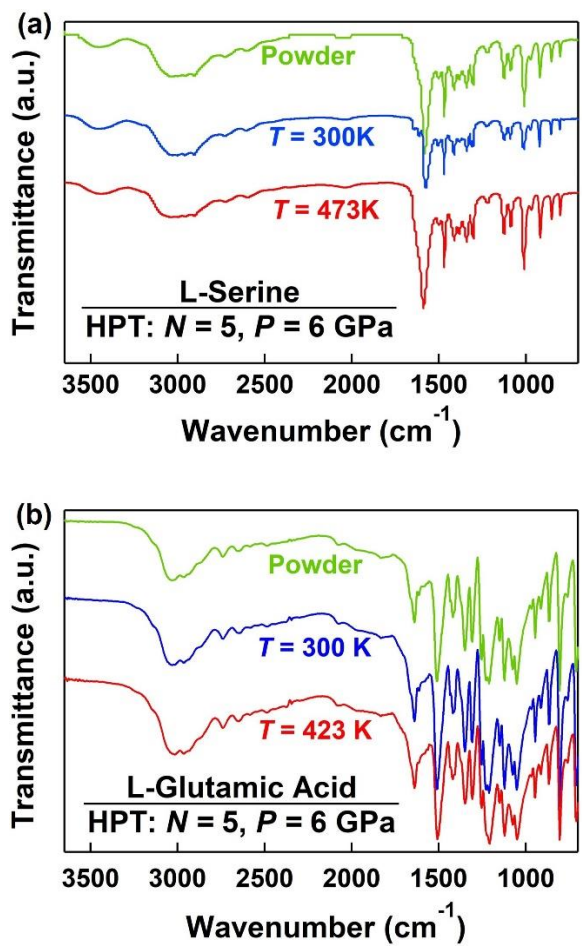

Figure 4. ATR-FTIR spectra for (a) *L*-serine and (b) *L*-glutamic acid before and after HPT treatment at room temperature and 423 K.

Further examination of chemical reactions by (a, b) $^1$H and (c, d) $^{13}$C NMR spectra is shown in Fig. 5 for (a, c) *L*-serine and (b, d) *L*-glutamic acid. NMR spectra are consistent with the literature [53] and no change corresponding to chemical reactions is detected. It is then concluded that *L*-serine and *L*-glutamic acid are quite resistant to chemical reactions during HPT processing, unlike glycine which showed chemical decomposition [10]. It should be noted that the current

NMR results cannot negate the possible formation of very small amounts of chemical reaction products due to the detection limits of NMR. To detect very small amounts of possible products, techniques based on chromatography and mass spectroscopy should be employed in future studies.

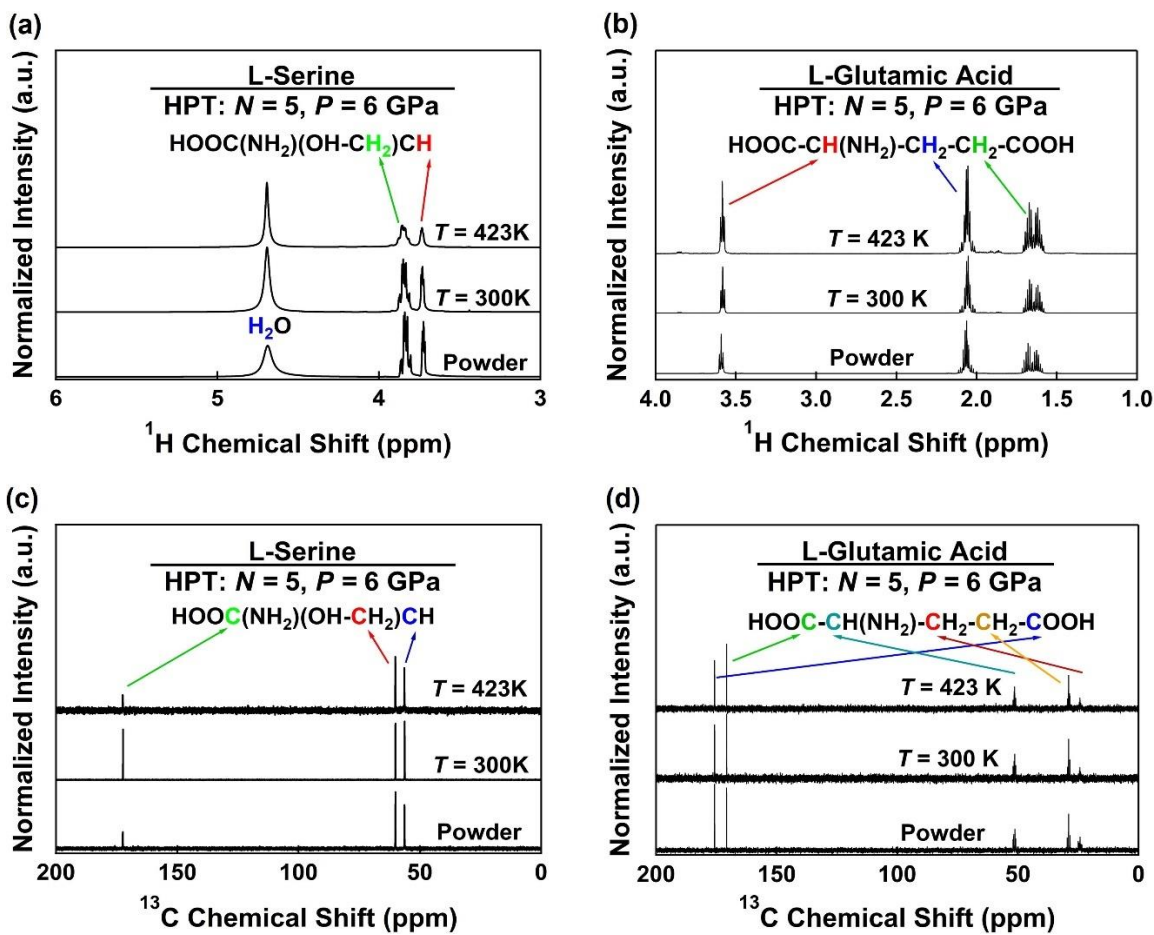

Figure 5. (a, b) $^1$H and (c, d) $^{13}$C NMR spectra for (a, c) *L*-serine and (b, d) *L*-glutamic acid before and after HPT treatment at room temperature and 423 K.

The mechanical behavior of amino acids under severe plastic deformation by HPT processing is shown in Fig. 6 as shear stress-strain curves. Since shear stress is estimated from torque measurement under high pressure, it is significantly overestimated due to the formation of a thin flash of material in the contact region of the upper and lower HPT anvils which experiences high friction forces. To have a better idea about the level of strength of these two amino acids, shear stress-strain curves for pure aluminum (99.99%) and copper (99.99%) were achieved under similar conditions. Shear strength for both amino acids increases with increasing strain at the early stages of straining but saturates at large strains. Such saturation of mechanical properties can be also observed for pure aluminum and copper in Fig. 6 and for many other metallic materials [12,13] and ceramics [30] in the literature. The steady-state stress for these two amino acids is higher than pure aluminum with a Vickers hardness of 310 MPa, but it is comparable to pure copper with a Vickers hardness of 1300 MPa. The occurrence of a steady state is due to a balance between hardening phenomena (crystal defect generation and grain refinement) and softening phenomena such as dynamic recovery [28], dynamic recrystallization [54] and grain boundary migration [14].

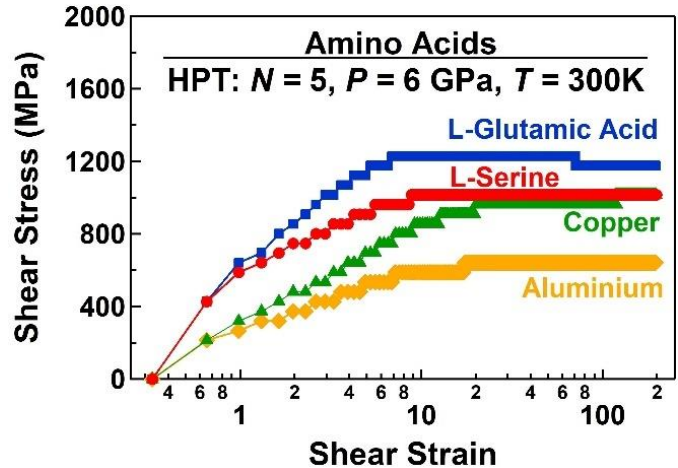

Figure 6. Shear stress versus shear strain curves for *L*-serine and *L*-glutamic acid processed by HPT at room temperature in comparison with pure aluminum (99.99%) and copper (99.99%) processed under the same conditions.

**Discussion**

The current study reports the mechanochemical behavior of two amino acids under high pressure and strain to simulate their behaviors during impacts by small solar system bodies.

Although shock experiments are usually used for this purpose [5-9], the HPT process was used in this study for such simulations. Shock experiments have the advantage of simulating large strain rates during the impacts, but strain is significantly neglected in shock experiments. HPT has the benefit of considering the strain during the impacts, but the strain rate in available HPT machines is still smaller than the one during real impacts. In addition to the simulation of astronomical impacts, this study contributes to the field of severe plastic deformation because few studies are available about the chemical and mechanical behaviors of biomolecules after severe plastic deformation [10]. The chemical and mechanical behaviors of amino acids need to be further discussed in this section.

Regarding the chemical response of these two amino acids to HPT processing conditions, no phase transformations or chemical reactions are detected within the limits of employed characterization methods. This is consistent with the discovery of these amino acids in some meteorites [2-4]. Contrary to *L*-serine and *L*-glutamic acid, glycine - which is the simplest amino acid and is considered highly stable and non-reactive - partially decomposes to alcohol during HPT processing [10]. To have a comparison, it should be noted that the thermodynamics of materials under extreme conditions of pressure and strain are not similar to their thermodynamics under ambient conditions. High pressure combined with strain introduces various defects and enhances the internal energy of the system, leading to thermodynamic changes and the occurrence of transformations that are impossible or hidden under ambient conditions [18,55,56]. Therefore, a comparison between the thermodynamic stability of glycine, serine, and glutamic acid under astronomical impact should be conducted by future theoretical studies under extreme conditions of pressure and strain. The absence of polymerization of serine and glutamic acid to proteins in this study cannot still negate such a possibility in the early Earth conditions. A difference between real impacts in the early Earth and current HPT experiments is that high-purity and dried amino acids were used in this study, while amino acids in real impacts co-exist with minerals, chemicals, and water [1,6]. It was shown that water and aqueous alternation can influence the formation of organic compounds in meteorites [57]. Minerals can also act as catalysts for organic reactions, especially in the presence of high-temperature water [58]. Therefore, future HPT experiments should be conducted on various amino acids after the addition of minerals, water, or chemicals that existed in the early Erath conditions. Moreover, since XRD and Raman spectroscopy experiments cannot be conducted in situ during HPT processing, future experiments using rotational (shear) diamond anvil cells can be useful to clarify the structural evolution of these amino acids under high pressure [55,59].

Regarding the mechanical and microstructural behavior of crystalline amino acids under HPT conditions, there are significant similarities to metallic materials and ceramics processed by severe plastic deformation [15]. Grain refinement to the nanometer level, formation of vacancies, and increasing the shear stress to a steady state level are some similarities that were reported frequently in metallic materials [13-15] and occasionally in ceramics [30,41]. Although it was not possible to examine these amino acids by transmission electron microscopy due to fast damage by electron beam, XRD peak broadening suggests that dislocations are also generated in these amino acids. Formation of vacancies and dislocations and grain refinement are some strain-induced effects that can be observed in various kinds of HPT-processed crystalline materials regardless of their metallic or non-metallic structures [12,15]. The appearance of a steady state at large strains for shear strain confirms that a balance between the generation of defects (vacancy, dislocation and boundaries) and the annihilation of defects should happen at large strains for amino acids as well. The main mechanisms reported for achieving such a balance and the appearance of a steady state are dynamic recovery [28], dynamic recrystallization [54] and grain boundary migration [14].

When a metal has a low melting point such as pure aluminum, recovery and recrystallization occur more easily, resulting in larger grain sizes and lower hardness and shear strength levels [60,61]. In materials with covalent bonding such as silicon or ceramics, recovery and recrystallization can occur with more difficulty compared to materials with metallic bonding (even compared to high-melting-point metals), resulting in their smaller grain sizes and higher shear stress or hardness levels [30,61]. Although current amino acids have low melting points (< 500 K), their covalent bonding should be responsible for their small crystallite sizes close to 20 nm after HPT processing, and for their high shear strengths which are comparable to copper which is a metal with a moderate melting point of 1357 K. From the results presented in this study, one may expect that the evolution of microstructure in these amino acids should follow the same stages reported for metals: dislocation formation, accumulation and arrangement of dislocations, formation of high-angle grain boundaries by continuous or discontinuous dynamic recrystallization, and appearance of steady state by dynamic recovery [28], dynamic recrystallization [54] and grain boundary migration [14]. Therefore, despite significant differences between the atomic bond parameters of metals, ceramics, and crystalline organic materials, their overall behavior is reasonably similar during severe plastic deformation, a fact that was occasionally reported by Bridgman several decades ago [11,12,22].

**Conclusions**

High-pressure torsion, as a new platform for the simulation of impacts by small solar system bodies, was applied to *L*-serine and *L*-glutamic acid. No phase transformation, chemical reaction, or polymerization was detected in pure and dry amino acids after HPT processing. The evolution of microstructure (lattice defect and nanograin formation) and mechanical properties (hardening to steady-state levels) for amino acids were principally similar to metals and ceramics, suggesting the occurrence of the same mechanisms for microstructural evolution/saturation in metals, ceramics, and crystalline organic compounds.

**Acknowledgments**

The author JHJ acknowledges a scholarship from the Q-Energy Innovator Fellowship of Kyushu University. This This study is supported partly by a Grant-in-Aid from the Japan Society for the Promotion of Science (JP22K18737), and partly by the Japan Science and Technology Agency (JST), the Establishment of University Fellowships Towards the Creation of Science Technology Innovation (JPMJFS2132).

**Conflict of Interest**

The authors declare no conflict of interest.